\documentclass{icrc2009}

\usepackage{graphicx}   
\usepackage{caption}    
\usepackage{fixltx2e}
\usepackage{url}

\newcommand{\shorttitle}[1]%
{\markboth{Proceedings of the 31\MakeLowercase{$^{st}$} ICRC, {\L}\'{o}d\'{z} 2009}{#1} }
\newcommand{\etal}{\MakeLowercase{\textit{et al. }}} 

\begin{document}
\title{X- and gamma-ray studies of HESS J1731-347 coincident with a newly discovered SNR }

\author{\IEEEauthorblockN{F. Acero\IEEEauthorrefmark{1},
                           G. P\"uhlhofer\IEEEauthorrefmark{2},
			   D. Klochkov\IEEEauthorrefmark{2},
                           Nu. Komin\IEEEauthorrefmark{3}, 
                           Y. Gallant\IEEEauthorrefmark{1},
                           D. Horns\IEEEauthorrefmark{4} and
                           A. Santangelo\IEEEauthorrefmark{2}\\
for the H.E.S.S. collaboration}
                            \\
\IEEEauthorblockA{\IEEEauthorrefmark{1} 
LPTA, CNRS/IN2P3, Universit\'e Montpellier II, Place Eug\`ene
Bataillon, 34095 Montpellier Cedex 5, France }

\IEEEauthorblockA{\IEEEauthorrefmark{2} 
  Institut f\"ur Astronomie und Astrophysik, Universit\"at T\"ubingen, Sand 1, D 72076 T\"ubingen, Germany }

\IEEEauthorblockA{\IEEEauthorrefmark{3} 
 IRFU/DSM/CEA, CE Saclay, F-91191 Gif-sur-Yvette, Cedex, France  }

\IEEEauthorblockA{\IEEEauthorrefmark{4} 
 Universit\"at Hamburg, Institut f\"ur  Experimentalphysik, Luruper Chaussee 149, D 22761, Hamburg, Germany  }

  }

\shorttitle{Acero \etal X- and gamma-ray studies of HESS J1731-347 }
\maketitle

\begin{abstract}

In the survey of the Galactic plane conducted with H.E.S.S.,
many VHE gamma-ray sources were discovered for which no clear counterpart
at other wavelengths could be identified. HESS J1731-347 initially belonged
to this source class. Recently however, the new shell-type supernova remnant
(SNR) G353.6-0.7 was discovered in radio data, positionally coinciding with
the VHE source. We will present new X-ray observations that cover a fraction
of the VHE source, revealing nonthermal emission that most likely can be
interpreted as synchrotron emission from high-energy electrons. This,
along with a larger H.E.S.S. data set which comprises more than twice the
observation time used in the discovery paper, allows us to test whether the
VHE source may indeed be attributed to shell-type emission from that new
SNR. If true, this would make HESS J1731-347 a new object in the small but
growing class of non-thermal shell-type supernova remnants with VHE emission.

  \end{abstract}

\section{Introduction}

Despite the meanwhile comparatively large number of known VHE sources in our Galaxy ($>$ 50), the number of identified VHE {\em shell-type} supernova remnants (SNR) is still surprisingly small. All SNR that are confirmed or likely shell-type VHE emitters are young (few 100 to few 1000 years): RX\,J1713.7-3946, RX\,J0852.0-4622, RCW\,86, SN\,1006, and Cas\,A\footnote{Cas\,A is however unresolved in TeV $\gamma$-rays.}, see \cite{horns} for a review. 

Many of the Galactic VHE sources found in surveys of the Galactic plane remain unidentified to date. Most of the sources are extended beyond the point spread function (PSF) of current VHE instruments ($\sim 0.1^{\circ}$). The largest number of successful identifications so far can be attributed to VHE pulsar wind nebulae (PWN). This comes slightly as a surprise because most VHE PWNe are spatially more extended than their counterparts (the X-ray PWN or the energetic radio pulsar driving the wind), rendering an identification through morphological correspondence often rather difficult. 

The first convincing candidate for a previously unidentified VHE source for which a firm identification with a {\em shell-type} SNR is within reach is HESS\,J1731-347. The radio SNR shell G353.6-0.7 that was discovered in spatial coincidence with the VHE source \cite{tian} has a diameter of nearly $0.5^{\circ}$, which allows -- given the brightness of the source and the VHE PSF -- for a morphological comparison of the VHE source with the radio shell. Moreover, at least up to the current date no radio pulsar or X-ray PWN candidate was found that might serve as alternative counterpart. This should be compared to other unidentified VHE sources like HESS\,J1813-178 or HESS\,J1640-465, which are also in spatial coincidence to radio SNR shells, but for which a morphological identification with the radio shells is not possible, and for which in addition also plausible PWN scenarios exist.

Given the only recent discovery of G353.6-0.7, little is known about its age and distance. \cite{tian} suggested a distance of 3.5 kpc and an age of 27\,kyears, based on assumptions which are still to be verified. Here, we report on investigations of X-ray and $^{12}$CO   line observations which help to constrain the SNR parameters. The X-ray observations with XMM-Newton, Chandra, and Suzaku confirm the suggestive X-ray counterpart found in archival ROSAT data. Moreover, we present new H.E.S.S. data that allow us to investigate the compatibility of the VHE source with the radio shell morphology.

\section{Non-thermal X-ray emission with shell-type morphology}
\label{sec:gradient}

\begin{figure*}
   \centering

\begin{tabular}{cc}
	\includegraphics[angle=-90,width=7.cm]{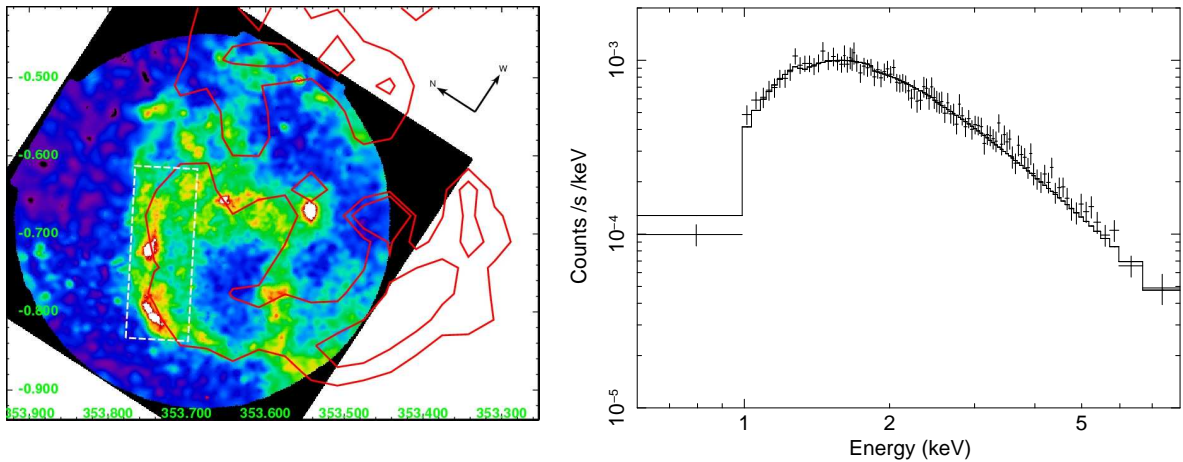}  &

	\includegraphics[width=6.cm,angle=-90]{icrc0678_fig02.ps} 
\end{tabular}
  
   \caption{ \textit{Left :} \textit{XMM-Newton} image of the MOS1 and MOS2 instrument in the 0.5 to 4.5 keV energy band. The image is exposure corrected and is adaptively smoothed to obtain a Signal/Noise ratio of 10. The radio contours from ATCA South Galactic Plane Survey are overlaid in red.
\textit{Right :} MOS 2 spectrum extracted from the dashed white region shown in the X-ray image.
 The spectrum is unfolded and the best-fit absorbed powerlaw model is shown. }
   \label{fig:spec}
\end{figure*}

The North-Eastern part of HESS\,J1731-347 was observed with Suzaku (20 ksec), XMM-Newton (23 ksec, see Fig.\,\ref{fig:spec} left panel), and Chandra (30 ksec). In all observations, a pointlike source at the center of the VHE source was detected, which we describe in more detail in the next section. Besides a few further, weak point sources, the X-ray emission is characterized by extended emission which is concentrated in arc-like features, similar to broken shell emission from shell-type SNRs. Some of the arcs follow the shell as outlined by the radio emission (see Fig\,\ref{fig:spec}, left panel). Some of the structures could hint at an additional, smaller shell, but might also come from irregular SNR expansion in an inhomogeneous and/or very dense medium. More detailed studies are ongoing; for the purpose of this paper, we assume for now that the diffuse X-ray emission -- at least from the NE arc -- can be attributed to G353.6-0.7.

The energy spectrum of the diffuse X-ray emission can be characterized by absorbed single power law models (see Fig. \ref{fig:spec}, right panel), with however different absorption column $N_{\mathrm{H}}$ and power law index $\Gamma$ at different positions in the field of view (FoV). The power law index is in most locations in the range $\Gamma = 2.1 .. 2.5$; the spectrum can well be explained by synchrotron emission from TeV electrons. 

\begin{figure*}
   \centering

\begin{tabular}{rl}
	\includegraphics[width=8.5cm]{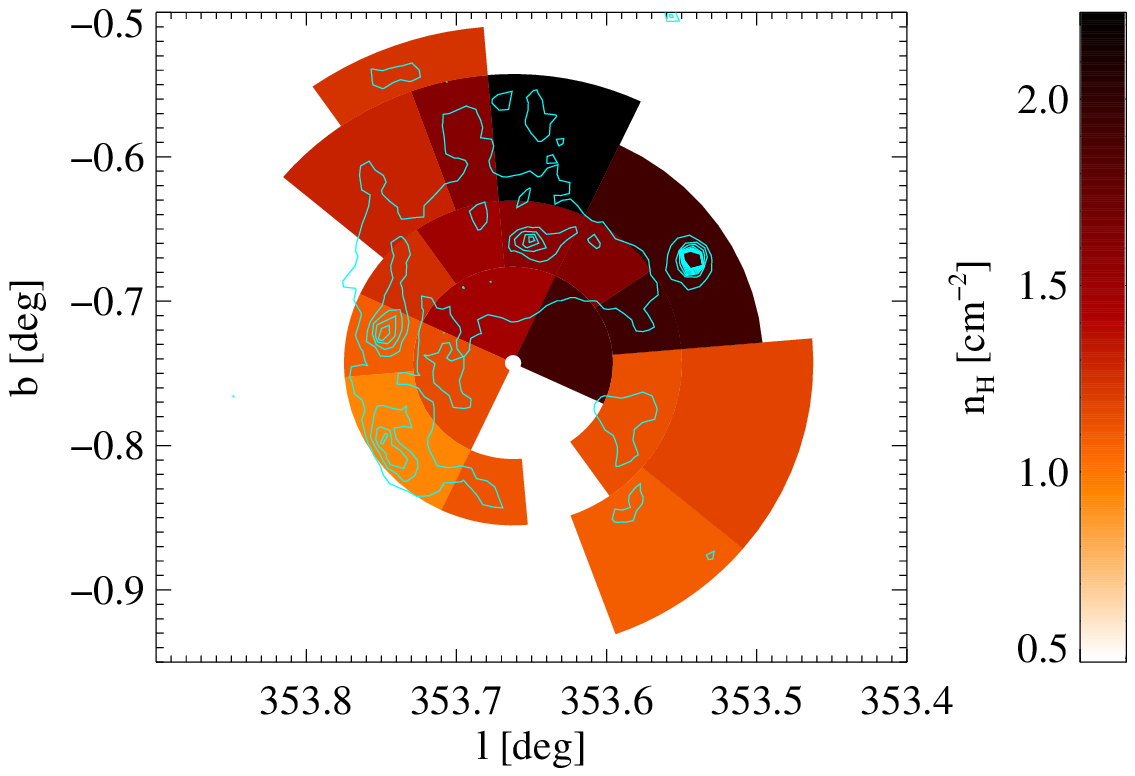} &  \vspace{-0.cm}
	\includegraphics[width=7.5cm]{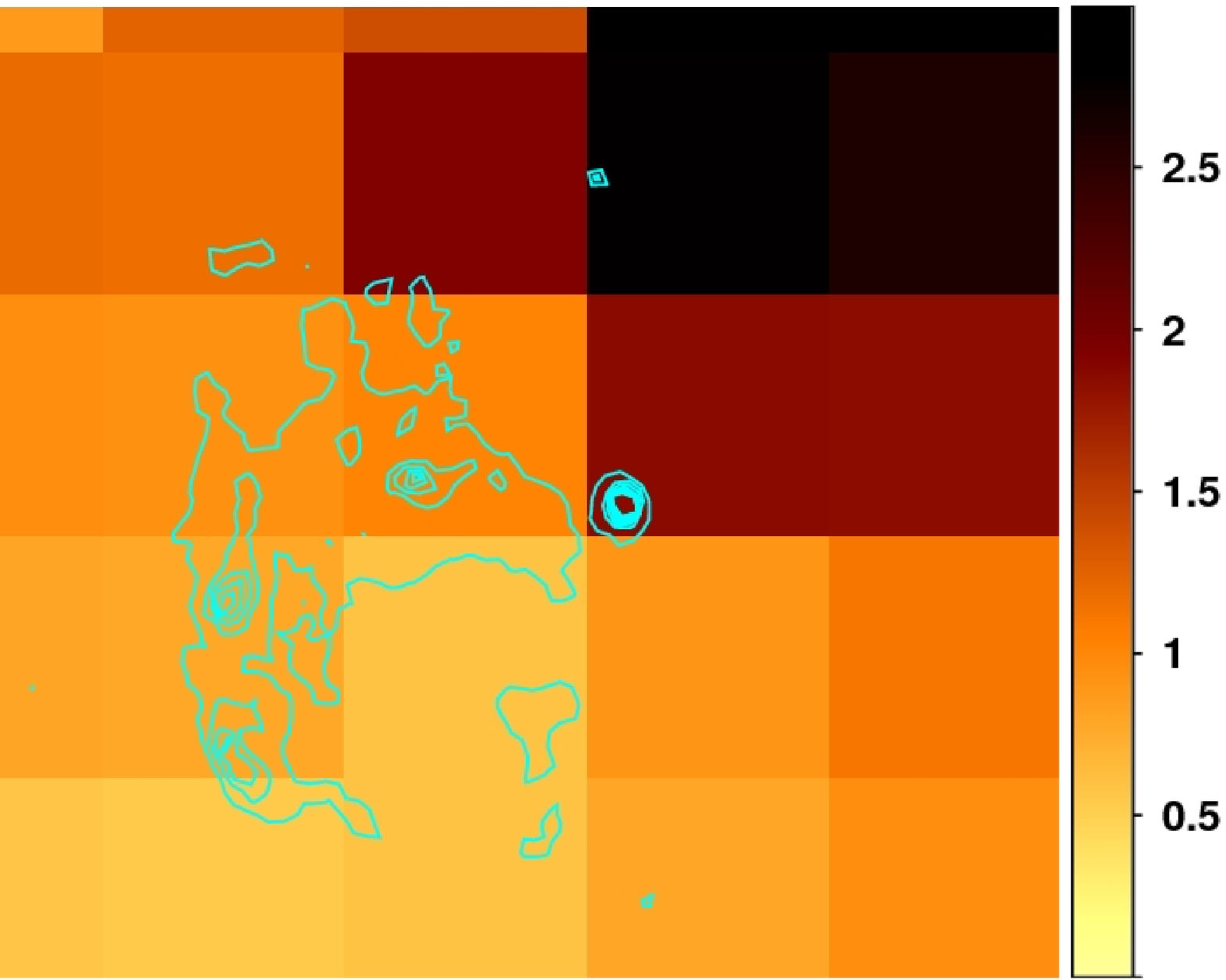}   
 
\end{tabular}
  
   \caption{ \textit{Left :}  Absorption map derived from an X-ray spectral analysis in each region. A gradient of absorption towards the galactic plane is seen. 
   \textit{Right :}  Absorption map toward the direction of the remnant derived from $^{12}$CO observations (integrated from LSR velocities between 0 and -17 km/s).
   X-ray contours obtained from Fig. \ref{fig:spec} are overlaid on the two maps. 
   The absorption is in units of $10^{22}$ cm$^{-2}$.}

\label{fig:gradient}
\end{figure*}

Under this assumption of a pure power-law hypothesis, there is significant variation of the absorption column, with the highest value ($N_{\mathrm{H}} = 1.7 \times 10^{22}$cm$^{-2}$) towards the NW of the FoV, see left panel of Fig.\,\ref{fig:gradient}. The gradient from the SE ($N_{\mathrm{H}} = 1.0 \times 10^{22}$cm$^{-2}$) towards the NW is much larger than what could be expected on average from the Galactic plane, and is therefore well compatible with the picture that the SNR is expanding into a dense molecular cloud, as discussed later in this paper.

\section{A central compact object at the center of G353.6-0.7 / HESS\,J1731-347}

At the center of G353.6-0.7 shell lies a compact (unresolved) X-ray source. Its XMM-Newton energy spectrum can
well be described by an absorbed blackbody spectrum with $k_{\mathrm{e}}T=0.5\,\mathrm{keV}$. There is no strong evidence of a power-law tail component in the spectrum. Assuming a pure blackbody spectrum, the absorption column can be well constrained to $N_{\mathrm{H}}=1.5 \times 10^{22}\mathrm{cm^{-2}}$. Adopting the lower limit distance estimate of 3.5 \,kpc derived later in Section \ref{sec:distance}, the luminosity of the source would amount to $L_{0.5-10\,\mathrm{keV}}  \geq 3.9 \times 10^{34}\mathrm{erg\,s^{-1}}$, corresponding to a size of the emission area of $R \geq $ 2\,km. 

The source flux is consistent with being constant, both from comparing the integrated fluxes of the three X-ray exposures as well as from an investigation of the individual lightcurves. A search in the XMM-Newton EPIC PN data did not reveal any evidence for  pulsations either. We modeled the Chandra PSF with Chaser and MARX at the reconstructed point source position including the effects of pileup, and compared it to the Chandra data, and found no evidence for a source extension.

All these properties are highly reminiscent of Central Compact Objects (CCOs) found in several other supernova shells (e.g. \cite{pavlov,gotthelf}). Based on its position at the geometrical center of the radio/VHE shell, we therefore associate the compact X-ray source 
with the SNR G\,353.6-0.7 and tentatively identify it as a member of the CCO class.

The blackbody emission area should not exceed the neutron star surface area, therefore one can in principle derive constraints on the neutron star's (and therefore the associated SNR's) distance, independent of other methods such as used by \cite{tian} or as described later in this paper. However, only distances larger than $\sim 20\,\mathrm{kpc}$ (90\% C.L.) can be excluded using this condition.\footnote{This estimate does not yet take into account the modification of the emission spectrum by the neutron star's atmosphere.} On the other hand, small distances (corresponding to small emission areas) seem disfavoured because of the lack of pulsations found in the X-ray spectra of the CCO.

\section{H.E.S.S observations and results}
\label{sec:gamma}

The gamma-ray source  HESS J1731-347 was first discovered in the survey of the Galactic plane 
and presented as an extended source in \cite{aha08}.
In this first data set,  14 hours of observation time was available. As of today, 
 the total H.E.S.S. observation time for this object
reaches about 30 hours after data quality cuts. This new data set,
 was analysed using the Combined Model-Hillas analysis \cite{denaurois}. 
By combining the discriminating parameters from two independent analyses, this method leads
to a better sensitivity than each method separately.
The maps were generated using the so-called ``hard-cuts'' to achieve good spatial resolution
 (r$_{68\%}$ of the  PSF  : 0.06$^{\circ}$).

\begin{figure}
   \centering

	\includegraphics[width=7.9cm]{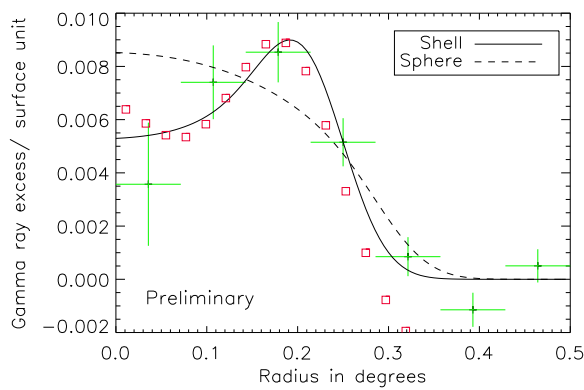}   
  
   \caption{ The gamma-ray excess (preliminary) and radio radial profiles are shown with  green crosses and red squares respectively. 
The best-fit smoothed projected emitting shell and sphere model are also shown. Both radial profiles 
are centered on the Compact Central Object ($l=353.54^{\circ}$, $b=-0.67^{\circ}$).   }

\label{fig:profile}
\end{figure}

The preliminary results are suggestive of a shell-like morphology in spatial coincidence 
with the shell seen in radio. To compare the morphology in the two energy bands we extracted radial 
profiles from the uncorrelated gamma-ray excess map and from the ATCA SGPS data (excluding point sources).
The radio profile is smoothed to match the H.E.S.S. spatial resolution and scaled  by a normalisation factor
calculated as the ratio of the total number of $\gamma$-rays excess over the total radio flux 
on the whole remnant. \\
The resulting profiles, presented in Fig. \ref{fig:profile}, show an extended emission in $\gamma$-rays
similar to the one seen in radio. 
To test  further the hypothesis of a shell morphology in gamma-rays we compared the radial profile
with  a sphere and a shell model (in dashed and solid line respectively in Fig. \ref{fig:profile}). 
The first model is a uniformly emitting sphere of variable radius, projected on the sky and then folded with  the H.E.S.S. PSF. The shell model consists of an emitting shell of variable radius  and thickness projected on the sky and then smoothed. In both cases the center was fixed to the position of the CCO
  ($l=353.54^{\circ}$, $b=-0.67^{\circ}$).
Even though the best-fit  favours marginally the shell model, the sphere model  is 
only ruled out at 2.1$\sigma$ ($\chi^{2}$/dof = 5.7/4  and $\chi^{2}$/dof = 11.6/5  
for the shell and sphere model respectively). Therefore any
conclusions on the $\gamma$-ray morphology will have to await further observations.  
In the case of a shell model, the best-fit  radius is 0.25$^{\circ}\pm$0.02$^{\circ}$ for a thickness 
of 0.05$^{\circ}\pm$0.03$^{\circ}$.

\section{Distance to the remnant}
\label{sec:distance}

The X-ray spectral analysis revealed a gradient of absorption towards the galactic plane.
A possible picture is the presence of a cloud overlapping a part of the remnant
 in the line of sight. This can be tested using the CfA $^{12}$CO survey data \cite{dame01}.
A CO spectrum towards the highest absorption region derived from the X-ray data 
($N_{\mathrm H}$=1.7$\times 10^{22}$ cm$^{-2}$ around $l=353.65^{\circ}$, $b=-0.55^{\circ}$)
exhibits two peaks of CO emission at \textit{Local standard of rest} (LSR) radial velocities 
of -17 km/s and -81 km/s. 
Fig. \ref{fig:gradient} (Right panel) shows that the cumulative  absorbing column densities 
derived from CO observations (using a CO-to-H$_{2}$ mass conversion factor of 
$2.5\times10^{20}$ cm$^{-2}$ K$^{-1}$ km$^{-1}$ s, \cite{dame01})
match well those derived from the X-ray analysis when integrating  LSR velocities from 
0 to -17 km/s.\\ 
Using a circular  galactic rotation model \cite{fich89}, the 
nearest distance corresponding to the LSR velocity of -17 km/s is 3.5 kpc. 
This distance to the cloud in the foreground sets a lower limit for the distance of the remnant.
Using the angular size derived from the shell model (see Sect. \ref{sec:gamma}) the radius of 
the remnant is $\geq$ 15 pc. \\ 
There are no estimation of the ambient medium density yet but for such a radius 
 the remnant is likely to be old even if it evolves in a  low density ambient medium : using Sedov solutions  $t_{\rm SNR} \simeq 4800 (\frac{n_{0}}{0.1 \, {\rm cm }^{-3}})^{1/2}$ yrs for an energy of the explosion of $10^{51}$ ergs.

\section{Summary}

The X-ray observations, that cover only part of the radio SNR G353.6-0.7,
 have revealed an interesting shell-like structure.
The X-ray emission from this shell is non-thermal and can be interpreted as synchrotron from accelerated high energy electrons. The 
spectrum of an X-ray point-like source, lying at the geometrical center of the radio SNR, is compatible
with a blackbody spectrum indicating that the object could be the CCO of the SNR.
The $\gamma$-ray source HESS J1731-347 is in spatial coincidence  with the SNR and shows an 
extension compatible with the one seen in radio. The $\gamma$-ray radial profile shows a hint
for a shell morphology that is however not statistically significant with current data. 
Based on CO observations and $N_{H}$ obtained from X-rays, we derive a lower limit to the distance of the X-ray emitting shell of  3.5 kpc. 
If the morphology in $\gamma$-rays is confirmed with further observations this will make of HESS J1731-347
the most distant SNR with a spatially resolved $\gamma$-ray shell.

\begin{center}
\begin{small}ACKNOWLEDGMENTS          
\end{small}         
\end{center}

The support of the Namibian authorities and of the University of Namibia in facilitating the construction
and operation of H.E.S.S. is gratefully acknowledged, as is the support by the German Ministry for Education and Research (BMBF), the Max Planck Society, the French Ministry for Research, the CNRS-IN2P3 and the Astroparticle Interdisciplinary Programme of the CNRS, the U.K. Science and Technology Facilities Council (STFC), the IPNP of the Charles University, the Polish Ministry of Science and Higher Education, the South African Department of Science and Technology and National Research Foundation, and by the University of Namibia. We appreciate the excellent work of the technical support staff in Berlin, Durham, Hamburg, Heidelberg, Palaiseau, Paris, Saclay, and in Namibia in the
construction and operation of the equipment.

\end{document}